\documentclass{ws-procs9x6-cpt22}
\begin{document}

\newcommand{\refeq}[1]{(\ref{#1})}
\def\etal {{\it et al.}}

\title{The unambiguous Lorentz-violating induced terms}

\author{A.R.\ Vieira,$^1$}

\address{$^1$ Universidade Federal do Tri\^angulo Mineiro - Campus Iturama,\\
Iturama, MG 38280-000, Brasil}



\begin{abstract}
In this work, we show that surface terms, which map dependence on 
regularization, can be fixed requiring momentum routing invariance of
tadpoles or diagrams with more external legs. This condition makes
the Lorentz-violating terms induced by quantum corrections determined
and uniques.
\end{abstract}


Lorentz and CPT violating operators reappear in quantum corrections and are called induced terms. If the coupling generates a loop with a 
finite integral, the induced term is well defined \cite{Petrov}. On the other hand, if the loop correction is divergent, it requires a regularization and a renormalization scheme.
The choice of the former might lead to spurious symmetry breaking terms and to ambiguous induced terms. Since the finite pieces of amplitudes of the Standard Model Extension
(SME) are in principle observable, they should not depend on this choice. However, symmetries of the built model are used to decide what kind of induced term is allowed and we 
can also find out if spurious terms appeared because of the choice of regularization \cite{brett}.

  
The question if a symmetry of the classical theory is also a symmetry of the quantum theory is non-trivial. The breaking of a classical symmetry at the quantum level is called 
an anomaly and the ideal regularization should preserve all symmetries so that it can tell us if the anomaly is physical or spurious. In the SME, this 
question is even more interesting because Lorentz or CPT symmetries are broken at the classical theory by one of the SME coefficients but the same coefficient might not 
appear at the quantum level or it may be a correction to a tree level term from a different sector. For instance, the $b^{\mu}$ coefficient of the fermion sector appears as a 
correction to the CS-like term $(k_{AF})^{\mu}$ from the photon sector.

There is no such ideal regularization but the spurious breaking of the symmetries is usual related to the assumption of a regulator. If it is not assumed, the so called implicit
regularization \cite{Orimar}, the breaking of the symmetries is mapped on surface terms. These surface terms manifest themselves as differences between two infinities. They can 
be any number and depend on the regularization scheme. They are born zero in dimensional regularization and if explicit computed, with a hard cutoff for instance, furnish a 
finite or a divergent result. There are several examples where they show up, especially when computing the breaking of a classical symmetry by quantum corrections. Let is consider
the QED vacuum polarization tensor as an example:
\begin{eqnarray}
&\small  \Pi^{\mu\nu}(p)=\frac{4}{3}(p^2\eta^{\mu\nu}-p^{\mu}p^{\nu})I_{log}(m^2)-
4\upsilon_2 \eta^{\mu\nu} - \nonumber\\
& +\frac{4}{3}(p^2\eta^{\mu\nu}-p^{\mu}p^{\nu})\upsilon_0 - \frac{4}{3}(p^2\eta^{\mu\nu}+2p^{\mu}p^{\nu})(\xi_0 -2\upsilon_0 ) -\nonumber\\
&-\frac{8i}{(4\pi )^2}(p^2\eta^{\mu\nu}-p^{\mu}p^{\nu})\int^1_0 x(1-x) \log \frac{m^2-p^2 x(1-x)}{m^2},
\label{vacpol}
\end{eqnarray}
where $I_{log}(m^2)=\int \frac{d^4k}{(2\pi )^4} \frac{1}{(k^2-m^2)^{2}}$ is a logarithmic basic divergent integral, $\upsilon_2$ is a quadratic surface term, $\upsilon_0$ and 
$\xi_0$ are logarithmic surface terms. The basic divergent integrals are defined as 
\begin{equation}
\small I^{\mu_1 \cdots \mu_{2n}}_{log}(m^2)\equiv \int \frac{d^4k}{(2\pi )^4} \frac{k^{\mu_1}\cdots k^{\mu_{2n}}}{(k^2-m^2)^{2+n}}
\end{equation}
and
\begin{equation}
\small I^{\mu_1 \cdots \mu_{2n}}_{quad}(m^2)\equiv \int \frac{d^4k}{(2\pi )^4} \frac{k^{\mu_1}\cdots k^{\mu_{2n}}}{(k^2-m^2)^{1+n}},
\end{equation}
while the surface terms are defined according to the relations \footnote{The curly brackets stand for permutation of indices,
$A^{\{\mu\nu}B^{\alpha\beta\}}=A^{\mu\nu}B^{\alpha\beta}+A^{\mu\alpha}B^{\nu\beta}+A^{\mu\beta}B^{\nu\alpha}$, for instance.}
\begin{align}
&\small \upsilon_{2w}g^{\mu \nu}=  g^{\mu \nu}I_{2w}(m^2)-2(2-w)I^{\mu \nu}_{2w}(m^2),
\nonumber\\
&\small \xi_{2w}g^{\{{\mu \nu}} g^{{\alpha \beta}\}}=  g^{\{ \mu \nu} g^{ \alpha \beta \}}I_{2w}(m^2)
 -4(3-w)(2-w)I^{\mu \nu \alpha \beta }_{2w}(m^2), \nonumber\\
&\small \sigma_{2w} g^{\{\mu \nu} g^{ \alpha \beta} g^ {\gamma \delta \}}= g^{\{\mu \nu} g^{ \alpha \beta} g^ {\gamma \delta \}}I_{2w}(m^2)-\nonumber\\
&-8(4-w)(3-w)(2-w) I^{\mu \nu \alpha \beta \gamma \delta}_{2w}(m^2),
\label{ST}
\end{align}
where $2w$ is the degree of divergence and we substitute the subscripts $log$ and $quad$ by $0$ and $2$, respectively.

The Ward-Takahashi identity ($p_{\mu}\Pi^{\mu\nu}=0$) we get with equation (\ref{vacpol}) reveals the quadratic surface term $\upsilon_2$ is zero and the relation 
$\xi_0=2\upsilon_0$. Thus, it is not always possible to fix all the surface terms requiring a symmetry. Another example similar to this one is the induced CS-like term in the 
massless case. The result is $\Pi^{\mu\nu}_5(p)= 4i\upsilon_0 b_{\alpha}p_{\beta}\epsilon^{\nu\alpha\mu\beta}$ \cite{CSlike}, where we can see the logarithmic surface term 
$\upsilon_0$ makes the induced term undetermined. Also, the Ward-Takahashi identity is compatible with any value of this surface term.

\begin{figure}
\centering
\includegraphics[trim=0mm 0mm 0mm 0mm, clip, scale=0.06]{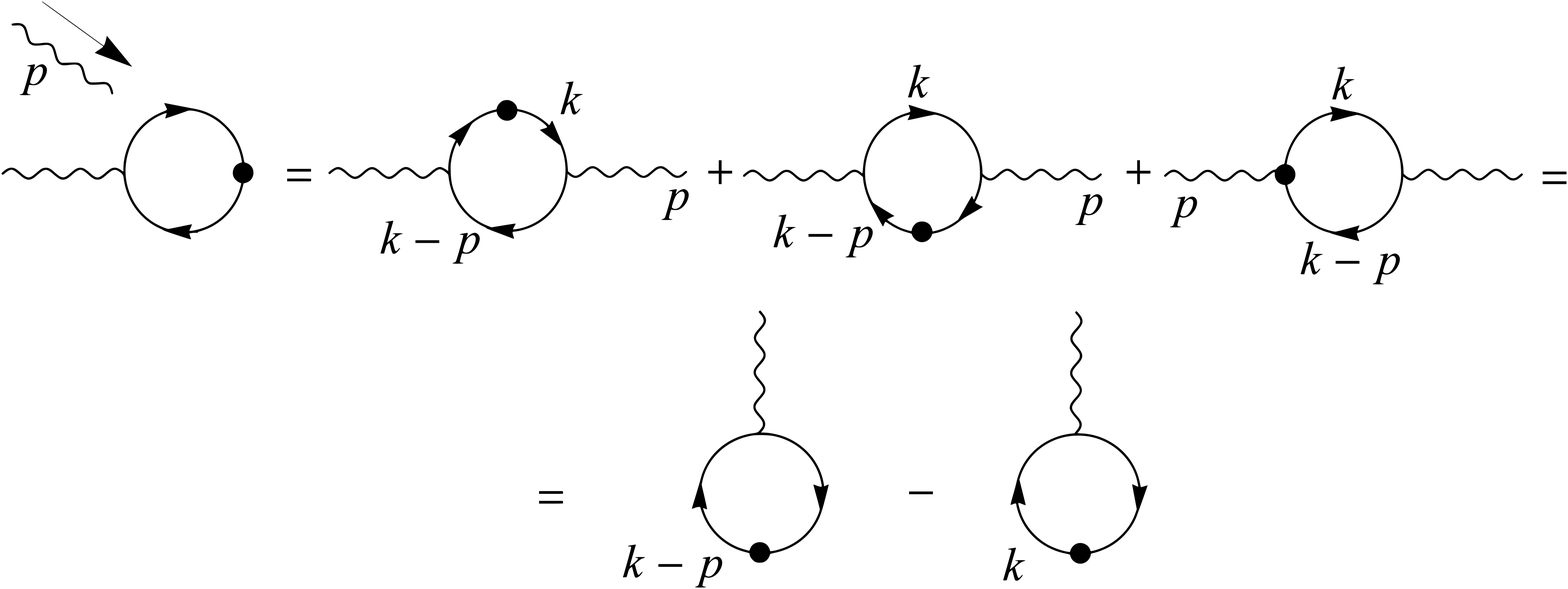}
\caption{The gauge and momentum routing invariance relation for a 2-point function diagram. The dot refers to the QED extension matrix $\Gamma^{\nu}=\gamma^{\nu}+
c^{\mu\nu}\gamma_{\mu}+d^{\mu\nu}\gamma_{5}\gamma_{\mu}+e^{\nu}+if^{\nu}\gamma_{5}+\frac{1}{2}g^{\lambda\mu\nu}\sigma_{\lambda\mu}$.}
\label{fig1}
\end{figure}

At the same time, there is a freedom of choosing the momentum routing in the internal lines of loop diagrams. If we set an arbitrary routing in these lines, constrained 
by momentum conservation at the vertices, this arbitrary routing is always accompanied by a surface term. For example, in the equation $\Pi^{\mu\nu}(p,l)=\Pi^{\mu\nu}(p,l')$, 
where $l$ and $l'$ are arbitrary routing. A textbook example where we can apply that is the axial anomaly. The usual procedure in its diagrammatic computation is choosing the 
routing so as to fulfill the gauge Ward-Takahashi identities and break the axial current by the known finite amount. However, this does not necessarily mean that the anomaly 
depends on the routing. It is possible to show in a diagrammatic computation that this result is valid for an arbitrary routing \cite{anomaly}. Furthermore, there is a 
one-to-one diagrammatic relation between gauge symmetry and momentum routing invariance in abelian and non-abelian theories. Figure \ref{fig1} shows the pictorial representation 
of a gauge Ward-Takahashi identity where we can see an external photon leg being inserted wherever is possible in a $2-$point function diagram, generating a difference between 
two $1-$point function diagrams with different routing. A similar relation can be obtained beyond one-loop and for diagrams with more external legs. 

The gauge and momentum routing invariance relation can be computed with implicit regularization and reveals a set of equations for the surface terms. In the 
QED extension this result yields for the $c^{\mu\nu}$ coefficient:
\begin{align}
&p_{\mu}\Pi^{\mu\nu}(p)=\tau^{\nu}(p)-\tau^{\nu}(0) = -16(3p^{\nu}c^{pp}+p^2c^{\nu p}+p^2c^{p \nu})(\xi_0 - \upsilon_0 )+\nonumber\\
&+4(p^2 me^{\nu}+2m p^{\nu}e\cdot p)(2\upsilon_0 -\xi_0 )+4\sigma_0 (2p^{\nu}c^{pp}+p^2c^{\nu p}+p^2c^{p \nu})+\nonumber\\
&+4 (\xi_2- 2\upsilon_2 )(c^{\nu p}+c^{p \nu})=0,
\label{eqGMRI}
\end{align}
where $\tau^{\nu}(p)$ is the tadpole for the full fermion propagator and $c^{p\nu}\equiv p_{\mu}c^{\mu\nu}$.

We see in equation (\ref{eqGMRI}) that the only possible solution is to make all surface terms zero. Thus, it is possible to fix 
the arbitrariness requiring momentum routing invariance and this automatically fulfill the gauge Ward-Takahashi identity. If the surface terms were 
computed with an explicit regulator like a hard cutoff, we would find spurious breaking of gauge invariance at one-loop. However, QED extension is gauge invariant
beyond tree level \cite{Tiago}. In this sense, we have determined induced terms. Some examples of the one-loop vacuum polarization tensor of the QED extension are 
\begin{align}
&\small \Pi^{\mu\nu}_{b}(p)=\left( 4ie^2\upsilon_0 +e^2\frac{m^2}{\pi^2}\iota_0\right) b_{\alpha}p_{\beta}\epsilon^{\alpha\beta\nu\mu}= e^2\frac{m^2}{\pi^2}\iota_0 b_{\alpha}p_{
\beta}\epsilon^{\alpha\beta\nu\mu} \nonumber\\
&\small \Pi^{\mu\nu}_{d}(p)=2ie^2(d_{\lambda\alpha}+d_{\alpha\lambda})\epsilon^{\lambda\mu\nu\beta}p_{\beta}p^{\alpha}(\upsilon_0-\xi_0 )=0 \nonumber\\
&\small \Pi^{\mu\nu}_{e,a}(p)=-4e^2( \eta^{\mu\nu}(m e\cdot p-a\cdot p)+p^{\mu}(m e^{\nu}-a^{\nu})+\nonumber\\
&+p^{\nu}(m e^{\mu}-a^{\mu}))(\xi_0-2\upsilon_0 )=0,
\label{UIT}
\end{align}
where $\iota_0=\int^1_0 dx \frac{(1-x)}{m^2-p^2 x(1-x)}$.

We see in eq. (\ref{UIT}) that we do not expect the induction of $d ^{\mu\nu}$, $e^{\mu}$ or $a^{\mu}$ coefficients and the mass term in the induced CS-like term avoids the 
infrared divergence. This is in agreement with recent results \cite{Weyl} where only $g^{\mu\nu\alpha}$ and $b^{\mu}$ coefficients appear in the low-energy limit of the one-loop 
vacuum polarization tensor.

\end{document}